\documentclass[journal]{IEEEtran}

\usepackage{cite}

\usepackage{color}

\usepackage{amsmath}

\usepackage{algorithm}
\usepackage{algorithmic}
\usepackage{bm}
\usepackage{latexsym}
\usepackage{amsthm}
\usepackage{url}
\usepackage{amsfonts}
\usepackage{amssymb}
\usepackage{indentfirst}
\usepackage{float}

\ifCLASSINFOpdf
   \usepackage[pdftex]{graphicx}
   \graphicspath{{../pdf/}{../jpeg/}}
   \DeclareGraphicsExtensions{.pdf,.jpeg,.png}
\else
   \usepackage[dvips]{graphicx}
   \graphicspath{{../eps/}}
   \DeclareGraphicsExtensions{.eps}
\fi
\usepackage{array}
\usepackage{multirow}

\ifCLASSOPTIONcompsoc
  \usepackage[caption=false,font=normalsize,labelfont=sf,textfont=sf]{subfig}
\else
  \usepackage[caption=false,font=footnotesize]{subfig}
\fi

\begin{document}
%
\title{Adaptive Convolutional Neural Network for Image Super-resolution}
%
%
%
\author{Ziang Wu,
        Jinwei Xie,
        Xuanyu Zhang,
        Tao Wang,
        Yongjun Zhang, 
        Qi Zhu, 
        Chunwei Tian\\
\thanks{This work was supported in part by the Basic and Applied Basic Research Foundation of Guangdong Province under grant 2025A1515011566, in part by Leading Talents in Gusu Innovation and Entrepreneurship under grant ZXL2023170, and in part by the Basic Research Programs of Taicang 2024 under grant TC2024JC32.
(Corresponding author: Chunwei Tian (Email:chunweitian@163.com) and Tao Wang (Email:twang@nwpu.edu.cn).)}

\thanks{Authors' Contact Information: Ziang Wu, School of Engineering, The Hong Kong University of Science and Technology, Hong Kong, China, ziang.wu@connect.ust.hk; Jinwei Xie, Macao Polytechnic University, Macao, China, xiejinwei8@gmail.com; Xuanyu Zhang, School of Software, Northwestern Polytechnical University, Xi’an, China, xuanyuzhang@mail.nwpu.edu.cn. Tao Wang, School of Computer Science, Northwestern Polytechnical University, Xi’an, China, twang@nwpu.edu.cn; Yongjun Zhang, College of Computer Science and Technology, Guizhou University, Guiyang, China, zyj6667@126.com; Qi Zhu, School of Artificial Intelligence, Nanjing University of Aeronautics and Astronautics, Nanjing, China, zhuqi@nuaa.edu.cn; Chunwei Tian, Shenzhen Research Institute of Northwestern Polytechnical University, Northwestern Polytechnical University, Shenzhen, China, Yangtze River Delta Research Institute, Northwestern Polytechnical University, Taicang, China, chunweitian@163.com.}}

\maketitle

\begin{abstract}
Convolutional neural networks can automatically learn features via deep network architectures and given input samples. However, the robustness of obtained models may face challenges in varying scenes. Bigger differences in network architecture are beneficial to extract more diversified structural information to strengthen the robustness of an obtained super-resolution model. In this paper, we proposed a adaptive convolutional neural network for image super-resolution (ADSRNet). To capture more information, ADSRNet is implemented by a heterogeneous parallel network. The upper network can enhance relation of context information, salient information relation of a kernel mapping and relations of shallow and deep layers to improve performance of image super-resolution. That can strengthen adaptability of an obtained super-resolution model for different scenes. The lower network utilizes a symmetric architecture to enhance relations of different layers to mine more structural information, which is complementary with a upper network for image super-resolution. The relevant experimental results show that the proposed ADSRNet is effective to deal with image resolving. Codes are obtained at https://github.com/hellloxiaotian/ADSRNet.
\end{abstract}


\section{Introduction}
\IEEEPARstart{S}{ingle} image super-resolution (SISR) techniques can obtain high-quality images from given low-resolution images (LR), according to solution of ill-posed inverse problems \cite{zha2019rank, zhou2012single}. In recent years, machine learning techniques have obtained huge achievements in various applications, i.e., disease diagnosis\cite{wang2023deep}, classification\cite{zhou2023information,zhu2022deep}, object recognition\cite{zhang2023deep}, multimodal data fusion\cite{zhang2020advances, zhou2024spatial}, Metaverse\cite{zhou2024personalized, zhou2023digital} and SISR \cite{zhou2023msra,amanatiadis2015accelerating}. Specifically,
deep learning techniques with end-to-end architectures have obtained better visual effect in SISR \cite{wanga2023hybrid}. Deep convolutional neural networks (DCNNs) use end-to-end architectures rather than manual setting parameters to obtain strong learning abilities to improve visual effects of SISR\cite{jiang2020hierarchical}.  Dong et al. designed 3-layer network through pixel transform  to convert a low-resolution (LR) image to a high-resolution (HR) image \cite{dong2015image}. Although this method can improve resolution of predicted images, it is limited between network depth and performance in SISR. To address this issue, scholars try to enlarge network depth to pursue improved performance in image super-resolution \cite{kim2016accurate}. Stacking small filters is used to achieve a very deep SR network \cite{kim2016accurate}. Additionally, a residual learning technique (RL) is employed on a deep network layer in a network to make a balance between SR performance and computational costs. A deeply-recursive convolutional network (DRCN) \cite{kim2016deeply} and deep recursive residual network (DRRN) \cite{tai2017image} can exploit recursive networks to decrease parameters of training a SR model. To reduce computational complexities, some up-sampling operations, such as transposed convolution \cite{dumoulin2016guide}, deconvolution and sub-pixel convolution \cite{shi2016real} are set on a deep network layer to amplify low-frequency information for constructing HR images. For instance, a fast SR convolutional neural network (FSRCNN) \cite{dong2016accelerating} directly inputted low-resolution images to a deep CNN to obtain low-frequency features and applies a upsampling operation to transform low-frequency features to high-frequency features and obtain HR images. To further improve SR performance, an enhanced deep SR network as well as EDSR used enhanced residual blocks to extract more structural information for image super-resolution \cite{lim2017enhanced}. Also, deleting many unnecessary batch normalization can improve effect of training efficiency a SR model \cite{lim2017enhanced}. Although these algorithms are effective for SISR, they did not automatically update parameters, according to different scenes. That may cause poor robustness of obtained SR models for complex scenes. we propose a heterogeneous dynamic network for SISR as well as ADSRNet. ADSRNet uses a heterogeneous parallel network to capture more information to improve performance of image SR in this paper. The upper network uses stacked heterogeneous blocks to extract more contexture information in image SR. Also, each heterogeneous block uses different components to achieve dynamically learning parameters to strengthen robustness of our ADSRNet for SISR, according to different scenes. Also, it can prevent long-term dependency problem. The lower network can use a symmetric architecture to enhance relationships of between different layers to obtain more complementary structural information. Quantitative and qualitative analysis show that the proposed method is a good tool for image super-resolution.

Our main contributions can be summarized as follows. 

(1)	A heterogeneous parallel network is used to facilitate complementary structural information to improve performance for image super-resolution in terms of contextual,  hierarchical and salient information. 

(2)	Dynamic convolutions are embedded into a convolutional neural network to strengthen robustness of obtained SISR model for complex scenes.

(3)	An enhanced residual architecture is designed to address long-term dependency for image super-resolution.

The remaining of this paper is listed. Section II describes related work. Section III illustrates the proposed method. Section IV collects massive comparative experimental results. And Section V summarizes our work.
\section{Related Work}
\subsection{Deep CNNs for Image Super-resolution}
Due to the shooting distance, captured images by cameras are unclear. Also, traditional image super-resolution methods need manual setting parameters and complex optimization parameters. To address this issue, deep learning techniques are extended for SISR \cite{tian2022generative, yu2023osrt}. For instance, a deep recursive residual network used local residual connection and unit to improve effects of image super-resolution \cite{tai2017image}. Alternatively, Yang et al. used recursive and gate units to transfer information shallow layers to deep layers to address long-term dependency problem \cite{tai2017memnet}.  To improve efficiency of image SR, given LR images are directly as an input of a convolutional neural network and high-quality images are constructed via an up-sampling operation in deep layer \cite{dong2016accelerating}.  For instance, Ahn et al. used group convolutions to implement a cascading residual CNN to extract more robustness low-frequency information and decrease the number of parameters without causing significantly performance loss in SISR \cite{ahn2018fast}. Tian et al. exploited symmetric network architecture to mine more accurate low-frequency information for image super-resolution \cite{tian2022heterogeneous}. Zhang et al. extended the depth of network and fully used hierarchical information to afford more low-frequency information for image SR \cite{zhang2018residual}. According to mentioned methods, we can see that deep CNNs are good tools to obtain clearer images and enhanced CNNs are effective for SISR. Thus, we enhanced a CNN to pursue better visual effects in SISR. Differing from previous research, we choose parallel and serial ways to facilitate heterogeneous information to better represent low-frequency structural information to improve visual effects of SISR.

\subsection{Dynamic convolution}
Existing CNNs can share parameters to extract useful features for better represent images. However, they may face challenges from varying input images in complex scenes \cite{chen2020dynamic}.  To address this issue, dynamic convolutions have been introduced \cite{chen2020dynamic}. That is, they can dynamically adjust parameters to adaptively learn models for image applications, according to different inputs. For instance, Wan et al. achieved a graph convolution by arbitrarily structuring non Euclidean data for hyperspectral image classification \cite{wan2019multiscale}. Alternatively, Ding et al. can atomically find valuable receptive fields of each target node to adaptively capture neighbor information in hyperspectral image classification \cite{ding2021adaptive}. To deal with native effects from incurring artifacts of textureless and edge regions, Duan et al. can dynamically generate kernels by region context from input images to better achieve image fusion \cite{duan2021dckn}. To adaptively adjust brightness of enhanced images, Dai et al. fused Taylor expansion and dynamic convolution into a Retinex to intelligently improve clarity of low-light images and degree of brightness flexibly \cite{dai2021single}.  To extract more context information, Hou et al. proposed dynamic hybrid gradient convolution and coordinate sensitive attention to enhance boundary information extraction for remote sensing image segmentation \cite{hou2022bsnet}. To find more high-frequency information, Shen et al. used a spatially enhanced kernel generation to dynamic learn high-frequency and multi-scale features to better achieve visual effect in image restoration \cite{shen2023adaptive}. Additionally, Tian et al. exploited relations of channels to extract richer low-frequency information for SISR \cite{tian2022heterogeneous}. Based on these examples, dynamic convolutions can dynamically learn parameters to improve robustness of an obtained CNN for SISR, according to different inputs. Thus, we use dynamic convolutions to guide a CNN to update parameters to strengthen adaptability in terms of recovering high-quality images for different scenes. Differing form previous research, a heterogeneous network architecture can facilitate more robustness features to generate high-quality images.
\section{Proposed Method}
\subsection{Network Architecture}
The proposed 18-layer ADSRNet contains two 16-layer parallel heterogeneous networks and a 2-layer construction block. The 16-layer parallel heterogeneous is composed of a 16-layer heterogeneous upper network and a symmetrical lower network. The heterogeneous upper network is composed of a Conv+ReLU (CRU) and five stacked heterogeneous blocks, which can extract more contexture information for image super-resolution. A CRU is a composite of a convolutional layer and a ReLU operation, which is used to extract non-linear information from given low-resolution images. Also, its input and output channel numbers are 3 and 64, respectively. Its kernel size is $3\times3$.These stacked heterogeneous blocks utilize different convolutional layers (i.e., dynamic and common convolutional layers) and ReLU to dynamically adjust parameters to obtain robust low-frequency features, according to low-resolution images of different inputs. To obtain complementary low-frequency features, a 16-layer symmetrical lower network is designed. It depends on a symmetrical architecture to enhance inter hierarchical connections to obtain more complementary structural information. Also, two sub-networks can interact information via a multiplication operation. A 2-layer construction block is used to transform low-frequency features to high-frequency features and construct predicted high-quality images. The process can be illustrated via the following formulates. 
\begin{equation}
    \label{eq1}
    \begin{array}{l}
    {I_S} = ADSRNet({I_L})\\
    {\rm{    }\quad} = CB{\rm{(}}HUNet({I_L}) \times SLNet({I_L}))\\
    {\rm{    }\quad} = CB{\rm{((}}5HB{\rm{(}}CR({I_L})){\rm{)}} \times SLNet({I_L}))\\
    {\rm{    }\quad} = CB({O_{HUNet}} \times {O_{SLNet}}),
    \end{array}
\end{equation}
where $ADSRNet$, $HUNet$ and $SLNet$ are used as ADSRNet, heterogeneous upper network and symmetrical lower network, respectively. $CR$ is a CRU. $5HB$ is five stacked heterogeneous blocks. $CB$ denotes a construction block. $\times$ represents multiplication operation. Parameters of obtained ADSRNet can be obtained via a loss function in Section III. B. ${O_{HUNet}}$ and ${O_{SLNet}}$ are outputs of the HUNet and SLNet, respectively. $I_L$ and $I_S$ represent the input low-resolution image and the output high-resolution image, respectively.
\begin{figure*}[!t]
\centering
\includegraphics[width=7.0in]{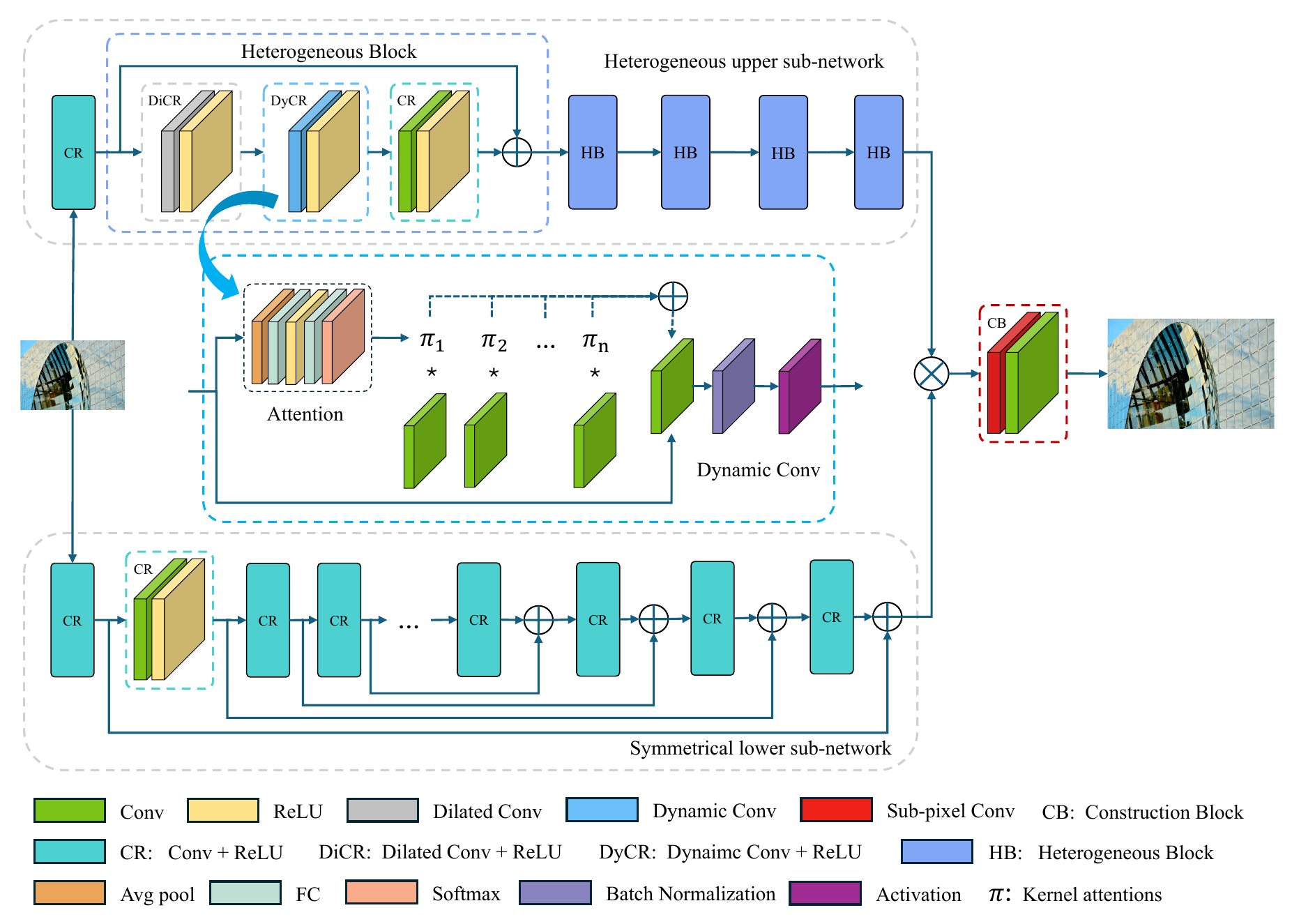}
\caption{Network architecture of the proposed ADSRNet.}
\label{fig_1}
\end{figure*}
\subsection{Loss Function}
ADSRNet choose mean absolute error (MAE) \cite{hui2018fast} as the loss function to obtain parameters. Work process of MAE in the ADSRNet can be represented as below. 
\begin{equation}
\label{eq2}
l(p) = {1 \mathord{\left/
 {\vphantom {1 T}} \right.
 \kern-\nulldelimiterspace} T}\sum\limits_{j = 1}^T {\left| {ADSRNet(I_L^j) - I_H^j} \right|},
\end{equation}
where $I_L^j$ and $I_H^j$ denote the $jth$ low- and high-resolution images, respectively. $ADSRNet$ demote function of ADSRNet, respectively. $T$ represents the number of low-resolution images. $l$ is a loss function of ADSRNet. Also, $p$ stands for parameters of the trained ADSRNet. Also, it can be optimized by the Adam optimizer \cite{kingma2014adam}. 
\subsection{Heterogeneous Block}
To improve the SISR effect and robustness of the network, heterogeneous blocks are conducted to dynamically adjust parameters to obtain robust low-frequency information, according to different input low-resolution images. Each heterogeneous block is composed of a dilated CRU, dynamic CRU and CRU, where dilated CRU represents a combination of a dilated convolution \cite{yu2015multi} and a ReLU \cite{krizhevsky2012imagenet}. That is used to capture more context information. A dynamic CRU is a combination of a dynamic convolution \cite{chen2020dynamic} and ReLU, where can adaptively learn parameters, according to different input information. To resolve long-term dependency, a RL operation is employed between an input of a dilated CRU and output of CRU. Mentioned convolutional kernels are $3\times3$. 
Input channel number and output channel number, i.e., dilated, dynamic and common convolutional layers are 64. Also, dilated factor is 2 in the dilated convolutional layers. The mentioned process can be conducted the following equation.
\begin{equation}
    \label{eq3}
    HB({O_t}) = CR(DyCR(DiCR({O_t}))) + {O_t},
\end{equation}
where ${O_t}$ is used to define an input of a heterogeneous block. $CR$ is a CRU. $DiCR$ stands for a dilated CRU. $DyCR$ represents a dynamic CRU. $+$ is used to express a RL operation also regarded to $\oplus$ in Fig. 1.
\subsection{Symmetrical Lower Sub-network}
To obtain complementary low-frequency information, a 16-layer symmetrical lower network is conducted. Each layer contains a CRU, where its channel number of input and output are 64 besides the first layer, its kernel is $3\times3$. Channel number of input and output from the first layer are set to 3 and 64, respectively. To enhance relations of different layers, residual learning operations are used to act between the 1st and 16th, the 2nd and 15th, 3nd and 14th, 4th and 13th, 5th and 12th, 6th and 11th, 7th and 10th, 8th and 9th layers to transfer obtained information of shallow layers to deep layers to prevent long-term dependency and obtain robust information for SISR. The procedure can be summarized as Eq. (4). 
\begin{equation}
    \label{eq4}
    \begin{array}{c}
    {O_{SLNet}} = CR(CR(...(CR(CR({O_8} + CR({O_8}))\\
     \qquad+{O_7}) + {O_6}) + ....) + {O_2}) + {O_1},
    \end{array}
\end{equation}
where ${O_i}$ denotes an output of the $ith$ layer and $i = 1,2,3...,8$. Also, ${O_i} = iCR({I_L})$, where $iCR$ stands for $i$ stacked CRU. $O_{SLNet}$ represent the output of SLNet, respectively. 
\subsection{Construction Block}
A 2-layer construction block is used to construct predicted HR images. It includes two steps. The first step is acted to a sub-pixel convolutional layer, which can use low-frequency features to obtain high-frequency features. Its input and output channel number are 128 and 64, respectively. The second phase only utilizes a convolutional layer (Conv) to construct predicted resolution images, where its input and output channel number are 64 and 3, respectively. Their kernel sizes are $3\times3$. These illustrations can be symbolled as Eq. (5).
\begin{equation}
    \label{eq5}
    \begin{array}{l}
    {O_S} = CB({O_{HUNet}} \times {O_{SLNet}})\\
    \quad\quad = C(Sub({O_{HUNet}} \times {O_{SLNet}})),
    \end{array}
\end{equation}
where $Sub$ and $C$ are functions of a sub-pixel convolutional layer and a convolutional layer, respectively. $CB$ denotes a construction block. $O_{HUNet}$ and $O_{SLNet}$ express the output of HUNet and SLNet.
\section{Experiments}
\subsection{Training Dataset}
We utilize the public DIV2K dataset\cite{agustsson2017ntire} as a training set to develop a ADSRNet for enhancing image super-resolution capabilities. Specifically, the DIV2K contains three parts, i.e., a training dataset containing 800 images, a validation dataset containing 100 images and a test dataset containing 100 images. To enlarge diversity of a training dataset, we merge an original dataset and a validation dataset to form a new training dataset, which includes 900 high-quality images and corresponding LR images of x2, x3 and x4. All training images are saved format of ‘.png’.
\subsection{Test Datasets}
To fairly test SR performance of our ADSRNet, four public SR datasets, i.e., Set5 containing 5 natural images \cite{bevilacqua2012low}, Set14 containing 14 natural images \cite{zeyde2012single}, BSD100 (B100) containing 100 natural images \cite{martin2001database} and Urban100 (U100) containing 100 containing \cite{huang2015single} are used to conduct experiments. These datasets include three different scales, i.e., x2, x3 and x4. They are saved as format of ‘.png’. 
\subsection{Implementation Details}
Original parameters of training a ADSRNet are $\beta1$ of 0.9, $\beta2$ of 0.999, epsilon of 1e-8, batch size of 64 and initial learning rate of 1e-4, which is halved every 300,000 steps. Also, it more parameters are the same as Ref. \cite{lim2017enhanced}.

The proposed ADSRNet is developed based on Pytorch 1.8.0 and Python 3.8. And all the experiments run on a workstation with Ubuntu 20.04, which equipped AMD EPYC 7502P CPU and four GPUs of Nvidia GeForce RTX 3090 with Nvidia CUDA 11.1, and cuDNN 8005. All experiments are trained via one 3090 GPU.
\subsection{Network Analysis}
We analyze architecture of designed ADSRNet containing a heterogeneous upper sub-network, symmetrical lower-network and construction block for image super-resolution, according to its rationality and validity. 

Heterogeneous upper sub-network: Most of existing SR models cannot adaptively learn parameters, according to different given low-resolution images \cite{xu2020unified}. Dynamic convolutions can automatically adjust parameters, according to different inputs \cite{chen2020dynamic}. In response to this motivation, we have developed a heterogeneous upper sub-network specifically for image super-resolution. A heterogeneous network architecture can facilitate more diversified structural information, which can provide complementary information to strengthen robustness of an obtained SISR model. Efficient super-resolution performance of proposed ADSRNet is implemented by enhancing relation of context information, salient information relation of a kernel mapping and relations of shallow and deep layers. That is, dilated convolutions are used to enhance relations of context information for SISR. Dynamic convolutions are used to enhance relations of salient information in a kernel mapping to improve adaptive ability of obtained ADSRNet. Also, residual learning technique is used to enhance relations of shallow and deep layers to prevent long-term dependency issue for SISR. More detailed design of the heterogeneous upper sub-network is given as follows. We design a main network, according to VGG \cite{simonyan2014very}. That is, six stacked CRU is used as a basic network to extract structural low-frequency information. A combination of six stacked CRU and a CB has obtained peak signal-to-noise ratio (PSNR) \cite{hore2010image} of 25.13dB and structural similarity (SSIM) \cite{hore2010image} of 0.7525 on U100 for x4, which shows effectiveness of six stacked CRU.  A CB is used to build HR images, which can be shown in latter section. To prevent long-term dependency problem, deep fusion idea \cite{jiang2020dual} is used in this paper. That is, each RL operation is used to act between an input and output of CRU besides the first CRU to transform low-frequency information from shallow layers to deep layers to pursue better performance on super-resolution. Its effectiveness can be shown by comparing ‘Heterogeneous upper sub-network without Dilated CRU, Dynamic CRU and a CB’ and ‘A combination of six stacked Conv’ in terms of PSNR and SSIM in TABLE I. To extract more context, each dilated CRU with dilated factor of 2 is used before each CRU besides the first CRU in heterogeneous upper sub-network. Its effectiveness can be proved by comparing ‘Heterogeneous upper sub-network without Dilated CRU, Dynamic CRU and a CB’ and ‘Heterogeneous upper sub-network without dynamic CRU and a CB’ in terms of PSNR and SSIM on U100 for x4. Specifically, five stacked blocks besides the first CRU in the heterogeneous upper sub-network (HUNet) are five heterogeneous blocks (HBs)  rationality and validity are shown in latter section. To make HUNet robust, each Dynamic CRU is set behind each Dilated CRU in each HB to dynamically adjust parameters, according to different inputs. Its better SR results can be found via comparing ‘Heterogeneous upper sub-network and a CB’ and ‘Heterogeneous upper sub-network without dynamic CRU and a CB’ in TABLE I, where effectiveness of dynamic convolutions is verified. It is known that difference of network architecture is bigger, its performance is better \cite{zhang2018residual}.  Motivated by that, we respectively use CRU rather than Dynamic CRU and Dilated CRU to conduct experiments. TABLE I shows that ‘Heterogeneous upper sub-network and a CB’ has obtained higher PSNR and SSIM values than that of ‘Heterogeneous upper sub-network with CRU rather than dilated CRU’ and ‘Heterogeneous upper sub-network with CRU rather than dynamic CRU’. 
This approach demonstrates the effectiveness of dynamic CRU and dilated CRU in HB, and highlights the advantages of various network architectures within HB for image super-resolution.

Symmetrical lower sub-network: 
To prevent poor representation of single network for image super-resolution, a symmetrical lower sub-network is designed, which can be used to assist an upper network to extract more complementary structural information to improve 
recovered high-resolution images. That is implemented by the 
following phases.

\begin{table}[!t]
\caption{SISR results of different methods on U100 for x4.\label{tab:table1}}
\centering
\scalebox{0.9}[0.95]{
\begin{tabular}{|c|l|l|}
\hline
\multicolumn{1}{|c|}{Scale} & \multicolumn{1}{c|}{Methods}                                                            & \multicolumn{1}{c|}{PSNR(dB)/SSIM} \\ \hline
\multirow{9}{*}{×4}         & A combination of six stacked CRU and a CB                                         & 25.13/0.7525                      \\ \cline{2-3} 
                            & Heterogeneous upper sub-network without Dilated                                         & \multirow{2}{*}{25.22/0.7559}  \\
                            & CRU, Dynamic CRU and a CB &                          \\ \cline{2-3}
                            & Heterogeneous upper sub-network without dynamic                       & \multirow{2}{*}{25.46/0.7649} \\
                            & CRU and a CB &                          \\ \cline{2-3}
                            & Heterogeneous upper sub-network without dilated                       & \multirow{2}{*}{25.61/0.7697} \\
                            & CRU and a CB &                          \\ \cline{2-3}
                            & Heterogeneous upper sub-network with CRU           & \multirow{2}{*}{25.74/0.7739} \\
                            & rather than dilated CRU &                          \\ \cline{2-3}
                            & Heterogeneous upper sub-network with CRU           & \multirow{2}{*}{25.68/0.7729} \\
                            & rather than dynamic CRU &                          \\ \cline{2-3}
                            & Heterogeneous upper sub-network and a CB                                                & 25.75/0.7740 \\ \cline{2-3} 
                            & ADSRNet without residual learning operations          & \multirow{2}{*}{25.99/0.7825} \\
                            & in symmetrical lower sub-network &                          \\ \cline{2-3}
                            & ADSRNet                                                                                 & 26.01/0.7827                       \\ \hline
\end{tabular}
}
\end{table}
\begin{table}[!t]
\caption{SISR results of different methods with on Set5 for three upscale factors.\label{tab:table2}}
\centering
\scalebox{0.82}[0.9]{
\begin{tabular}{|c|c|c|c|c|}
\hline
\multirow{2}{*}{Dataset} & \multirow{2}{*}{Methods} & ×2            & ×3            & ×4            \\ \cline{3-5} 
                         &                          & PSNR(dB)/SSIM & PSNR(dB)/SSIM & PSNR(dB)/SSIM \\ \hline
\multirow{28}{*}{Set5}   & Bicubic                  & 33.66/0.9299  & 30.39/0.8682  & 28.42/0.8104  \\ \cline{2-5} 
                         & SRCNN \cite{dong2015image}            & 36.66/0.9542  & 32.75/0.9090  & 30.48/0.8628  \\ \cline{2-5} 
                         & VDSR \cite{kim2016accurate}             & 37.53/0.9587  & 33.66/0.9213  & 31.35/0.8838  \\ \cline{2-5} 
                         & DRRN \cite{tai2017image}             & 37.74/0.9591  & 34.03/0.9244  & 31.68/0.8888  \\ \cline{2-5} 
                         & FSRCNN \cite{dong2016accelerating}           & 37.00/0.9558  & 33.16/0.9140  & 30.71/0.8657  \\ \cline{2-5} 
                         & CARN-M \cite{ahn2018fast}           & 37.53/0.9583  & 33.99/0.9236  & 31.92/0.8903  \\ \cline{2-5} 
                         & IDN \cite{hui2018fast}              & \textcolor{blue}{37.83/0.9600}  & 34.11/\textcolor{blue}{0.9253}  & 31.82/0.8903  \\ \cline{2-5} 
                         & A+ \cite{timofte2015a+}              & 36.54/0.9544  & 32.58/0.9088  & 30.28/0.8603  \\ \cline{2-5} 
                         & JOR \cite{dai2015jointly}             & 36.58/0.9543  & 32.55/0.9067  & 30.19/0.8563  \\ \cline{2-5} 
                         & RFL \cite{schulter2015fast}             & 36.54/0.9537  & 32.43/0.9057  & 30.14/0.8548  \\ \cline{2-5} 
                         & SelfEx \cite{huang2015single}          & 36.49/0.9537  & 32.58/0.9093  & 30.31/0.8619  \\ \cline{2-5} 
                         & CSCN \cite{wang2015deep}            & 36.93/0.9552  & 33.10/0.9144  & 30.86/0.8732  \\ \cline{2-5} 
                         & RED \cite{mao2016image}             & 37.56/0.9595  & 33.70/0.9222  & 31.33/0.8847  \\ \cline{2-5} 
                         & DnCNN \cite{zhang2012single}           & 37.58/0.9590  & 33.75/0.9222  & 31.40/0.8845  \\ \cline{2-5} 
                         & TNRD \cite{chen2016trainable}            & 36.86/0.9556  & 33.18/0.9152  & 30.85/0.8732  \\ \cline{2-5} 
                         & FDSR \cite{lu2018fast}            & 37.40/0.9513  & 33.68/0.9096  & 31.28/0.8658  \\ \cline{2-5} 
                         & RCN \cite{shi2017structure}             & 37.17/0.9583  & 33.45/0.9175  & 31.11/0.8736  \\ \cline{2-5} 
                         & DRCN \cite{kim2016deeply}            & 37.63/0.9588  & 33.82/0.9226  & 31.53/0.8854  \\ \cline{2-5} 
                         & LapSRN \cite{lai2017deep}             & 37.52/0.9590  & -  & 31.54/0.8850  \\ \cline{2-5}
                         &NDRCN \cite{cao2019new}             & 37.73/0.9596 & 33.90/0.9235 & 31.50/0.8859  \\ \cline{2-5}
                         & MemNet \cite{tai2017memnet}          & 37.78/0.9597  & 34.09/0.9248  & 31.74/0.8893  \\ \cline{2-5} 
                         & LESRCNN \cite{tian2020lightweight}         & 37.65/0.9586  & 33.93/0.9231  & 31.88/0.8903  \\ \cline{2-5} 
                         & LESRCNN-S \cite{tian2020lightweight}       & 37.57/0.9582  & 34.05/0.9238  & 31.88/0.8907  \\ \cline{2-5} 
                         & ScSR \cite{yang2010image}            & 35.78/0.9485  & 31.34/0.8869  & 29.07/0.8263  \\ \cline{2-5} 
                         & DSRCNN \cite{song2022dual}          & 37.73/0.9588  & \textcolor{blue}{34.17}/0.9247  & \textcolor{blue}{31.89/0.8909}  \\ \cline{2-5} 
                         & DAN \cite{huang2020unfolding}             & 37.34/0.9526  & 34.04/0.9199  & \textcolor{blue}{31.89}/0.8864  \\ \cline{2-5} 
                         & PGAN \cite{shi2023structure}            & -             & -             & 31.03/0.8798  \\ \cline{2-5} 
                         & ADSRNet   (Ours)         & \textcolor{red}{37.95/0.9604}  & \textcolor{red}{34.31/0.9263}  & \textcolor{red}{32.15/0.8927}  \\ \hline
\end{tabular}
}
\end{table}
\begin{table}[!t]
\caption{SISR results of different methods on Set14 for three upscale factors.\label{tab:table3}}
\centering
\scalebox{0.82}[0.9]{
\begin{tabular}{|c|c|c|c|c|}
\hline
\multirow{2}{*}{Dataset} & \multirow{2}{*}{Methods} & ×2            & ×3            & ×4            \\ \cline{3-5} 
                         &                          & PSNR(dB)/SSIM & PSNR(dB)/SSIM & PSNR(dB)/SSIM \\ \hline
\multirow{27}{*}{Set14}  & Bicubic                  & 30.24/0.8688  & 27.55/0.7742  & 26.00/0.7027  \\ \cline{2-5} 
                         & SRCNN \cite{dong2015image}            & 32.42/0.9063  & 29.28/0.8209  & 27.49/0.7503  \\ \cline{2-5} 
                         & VDSR \cite{kim2016accurate}             & 33.03/0.9124  & 29.77/0.8314  & 28.01/0.7674  \\ \cline{2-5} 
                         & DRRN \cite{tai2017image}             & 33.23/0.9136  & 29.96/0.8349  & 28.21/0.7720  \\ \cline{2-5} 
                         & FSRCNN \cite{dong2016accelerating}           & 32.63/0.9088  & 29.43/0.8242  & 27.59/0.7535  \\ \cline{2-5} 
                         & CARN-M \cite{ahn2018fast}           & 33.26/0.9141  & 30.08/0.8367  & 28.42/0.7762  \\ \cline{2-5} 
                         & IDN \cite{hui2018fast}              & 33.30/0.9148  & 29.99/0.8354  & 28.25/0.7730  \\ \cline{2-5} 
                         & A+ \cite{timofte2015a+}              & 32.28/0.9056  & 29.13/0.8188  & 27.32/0.7491  \\ \cline{2-5} 
                         & JOR \cite{dai2015jointly}             & 32.38/0.9063  & 29.19/0.8204  & 27.27/0.7479  \\ \cline{2-5} 
                         & RFL \cite{schulter2015fast}             & 32.26/0.9040  & 29.05/0.8164  & 27.24/0.7451  \\ \cline{2-5} 
                         & SelfEx \cite{huang2015single}          & 32.22/0.9034  & 29.16/0.8196  & 27.40/0.7518  \\ \cline{2-5} 
                         & CSCN \cite{wang2015deep}            & 32.56/0.9074  & 29.41/0.8238  & 27.64/0.7578  \\ \cline{2-5} 
                         & RED \cite{mao2016image}             & 32.81/0.9135  & 29.50/0.8334  & 27.72/0.7698  \\ \cline{2-5} 
                         & DnCNN \cite{zhang2012single}           & 33.03/0.9128  & 29.81/0.8321  & 28.04/0.7672  \\ \cline{2-5} 
                         & TNRD \cite{chen2016trainable}            & 32.51/0.9069  & 29.43/0.8232  & 27.66/0.7563  \\ \cline{2-5} 
                         & FDSR \cite{lu2018fast}            & 33.00/0.9042  & 29.61/0.8179  & 27.86/0.7500  \\ \cline{2-5} 
                         & RCN \cite{shi2017structure}             & 32.77/0.9109  & 29.63/0.8269  & 27.79/0.7594  \\ \cline{2-5} 
                         & DRCN \cite{kim2016deeply}            & 33.04/0.9118  & 29.76/0.8311  & 28.02/0.7670  \\ \cline{2-5} 
                         &LapSRN \cite{lai2017deep}             & 33.08/0.9130  & 29.63/0.8269 & 28.19/0.7720  \\ \cline{2-5}
                         & NDRCN \cite{cao2019new}             & 33.20/0.9141 & 29.88/0.8333 & 28.10/0.7697  \\ \cline{2-5}
                         & MemNet \cite{tai2017memnet}          & 33.28/0.9142  & 30.00/0.8350  & 28.26/0.7723  \\ \cline{2-5} 
                         & LESRCNN \cite{tian2020lightweight}         & 33.32/0.9148  & 30.12/0.8380  & 28.44/0.7772  \\ \cline{2-5} 
                         & LESRCNN-S \cite{tian2020lightweight}       & 33.30/0.9145  & 30.16/0.8384  & 28.43/0.7776  \\ \cline{2-5} 
                         & ScSR \cite{yang2010image}            & 31.64/0.8940  & 28.19/0.7977  & 26.40/0.7218  \\ \cline{2-5} 
                         & DSRCNN \cite{song2022dual} & \textcolor{blue}{33.43/0.9157}  & \textcolor{blue}{30.24/0.8402}  & \textcolor{blue}{28.46/0.7796}  \\ \cline{2-5} 
                         & DAN \cite{huang2020unfolding}             & 33.08/0.9041  & 30.09/0.8287  & 28.42/0.7687  \\ \cline{2-5} 
                         & PGAN \cite{shi2023structure}            & -             & -             & 27.75/0.8164  \\ \cline{2-5} 
                         & ADSRNet   (Ours)         & \textcolor{red}{33.52/0.9170}  & \textcolor{red}{30.27/0.8412}  & \textcolor{red}{28.56/0.7799}  \\ \hline
\end{tabular}
}
\end{table}
\begin{table}[!t]
\caption{SISR results of different methods on B100 for three upscale factors.\label{tab:table4}}
\centering
\scalebox{0.82}[0.9]{
\begin{tabular}{|c|c|c|c|c|}
\hline
\multirow{2}{*}{Dataset} & \multirow{2}{*}{Methods} & ×2            & ×3            & ×4            \\ \cline{3-5} 
                         &                          & PSNR(dB)/SSIM & PSNR(dB)/SSIM & PSNR(dB)/SSIM \\ \hline
\multirow{26}{*}{B100}   & Bicubic                  & 29.56/0.8431  & 27.21/0.7385  & 25.96/0.6675  \\ \cline{2-5} 
                         & SRCNN \cite{dong2015image}            & 31.36/0.8879  & 28.41/0.7863  & 26.90/0.7101  \\ \cline{2-5} 
                         & VDSR \cite{kim2016accurate}             & 31.90/0.8960  & 28.82/0.7976  & 27.29/0.7251  \\ \cline{2-5} 
                         & DRRN \cite{tai2017image}             & 32.05/0.8973  & 28.95/0.8004  & 27.38/0.7284  \\ \cline{2-5} 
                         & FSRCNN \cite{dong2016accelerating}           & 31.53/0.8920  & 28.53/0.7910  & 26.98/0.7150  \\ \cline{2-5} 
                         & CARN-M \cite{ahn2018fast}           & 31.92/0.8960  & 28.91/0.8000  & 27.44/0.7304  \\ \cline{2-5} 
                         & IDN \cite{hui2018fast}              & 32.08/0.8985  & 28.95/0.8013  & 27.41/0.7297  \\ \cline{2-5} 
                         & A+ \cite{timofte2015a+}              & 31.21/0.8863  & 28.29/0.7835  & 26.82/0.7087  \\ \cline{2-5} 
                         & JOR \cite{dai2015jointly}             & 31.22/0.8867  & 28.27/0.7837  & 26.79/0.7083  \\ \cline{2-5} 
                         & RFL \cite{schulter2015fast}             & 31.16/0.8840  & 28.22/0.7806  & 26.75/0.7054  \\ \cline{2-5} 
                         & SelfEx \cite{huang2015single}          & 31.18/0.8855  & 28.29/0.7840  & 26.84/0.7106  \\ \cline{2-5} 
                         & CSCN \cite{wang2015deep}            & 31.40/0.8884  & 28.50/0.7885  & 27.03/0.7161  \\ \cline{2-5} 
                         & RED \cite{mao2016image}             & 31.96/0.8972  & 28.88/0.7993  & 27.35/0.7276  \\ \cline{2-5} 
                         & DnCNN \cite{zhang2012single}           & 31.90/0.8961  & 28.85/0.7981  & 27.29/0.7253  \\ \cline{2-5} 
                         & TNRD \cite{chen2016trainable}            & 31.40/0.8878  & 28.50/0.7881  & 27.00/0.7140  \\ \cline{2-5} 
                         & FDSR \cite{lu2018fast}            & 31.87/0.8847  & 28.82/0.7797  & 27.31/0.7031  \\ \cline{2-5} 
                         & DRCN \cite{kim2016deeply}            & 31.85/0.8942  & 28.80/0.7963  & 27.23/0.7233  \\ \cline{2-5} 
                         & LapSRN \cite{lai2017deep}             & 31.80/0.8950  & -  & 27.32/0.7280  \\ \cline{2-5}
                         & NDRCN \cite{cao2019new}             & 32.00/0.8975 & 28.86/0.7991 & 27.30/0.7263  \\ \cline{2-5}
                         & MemNet \cite{tai2017memnet}          & 32.08/\textcolor{blue}{0.8978}  & 28.96/0.8001  & 27.40/0.7281  \\ \cline{2-5} 
                         & LESRCNN \cite{tian2020lightweight}         & 31.95/0.8964  & 28.91/0.8005  & 27.45/0.7313  \\ \cline{2-5} 
                         & LESRCNN-S \cite{tian2020lightweight}       & 31.95/0.8965  & 28.94/ 0.8012 & 27.47/0.7321  \\ \cline{2-5} 
                         & ScSR \cite{yang2010image}            & 30.77/0.8744  & 27.72/0.7647  & 26.61/0.6983  \\ \cline{2-5} 
                         & DSRCNN \cite{song2022dual} & \textcolor{blue}{32.05/0.8978}  & \textcolor{blue}{29.01/0.8029}  & 27.50/\textcolor{blue}{0.7341} \\ \cline{2-5}
                         & DAN \cite{huang2020unfolding}             & 31.76/0.8858  & 28.94/0.7919  & \textcolor{blue}{27.51}/0.7248  \\ \cline{2-5} 
                         & PGAN \cite{shi2023structure}            & -             & -             & 26.35/0.6926  \\ \cline{2-5} 
                         & ADSRNet   (Ours)         & \textcolor{red}{32.14/0.8994}  & \textcolor{red}{29.06/0.8048}  & \textcolor{red}{27.55/0.7357}  \\ \hline
\end{tabular}
}
\end{table}
\begin{table}[!t]
\caption{SISR results of different methods on U100 for three upscale factors.\label{tab:table5}}
\centering
\scalebox{0.8}[0.9]{
\begin{tabular}{|c|c|c|c|c|}
\hline
\multirow{2}{*}{Dataset} & \multirow{2}{*}{Methods} & ×2            & ×3            & ×4            \\ \cline{3-5} 
                         &                          & PSNR(dB)/SSIM & PSNR(dB)/SSIM & PSNR(dB)/SSIM \\ \hline
\multirow{25}{*}{U100}   & Bicubic                  & 26.88/0.8403  & 24.46/0.7349  & 23.14/0.6577  \\ \cline{2-5} 
                         & SRCNN \cite{dong2015image}            & 29.50/0.8946  & 26.24/0.7989  & 24.52/0.7221  \\ \cline{2-5} 
                         & VDSR \cite{kim2016accurate}             & 30.76/0.9140  & 27.14/0.8279  & 25.18/0.7524  \\ \cline{2-5} 
                         & DRRN \cite{tai2017image}             & 31.23/0.9188  & 27.53/0.8378  & 25.44/0.7638  \\ \cline{2-5} 
                         & FSRCNN \cite{dong2016accelerating}           & 29.88/0.9020  & 26.43/0.8080  & 24.62/0.7280  \\ \cline{2-5} 
                         & CARN-M \cite{ahn2018fast}           & 31.23/0.9193  & 27.55/0.8385  & 25.62/0.7694  \\ \cline{2-5} 
                         & IDN \cite{hui2018fast}              & 31.27/0.9196  & 27.42/0.8359  & 25.41/0.7632  \\ \cline{2-5} 
                         & A+ \cite{timofte2015a+}              & 29.20/0.8938  & 26.03/0.7973  & 24.32/0.7183  \\ \cline{2-5} 
                         & JOR \cite{dai2015jointly}             & 29.25/0.8951  & 25.97/0.7972  & 24.29/0.7181  \\ \cline{2-5} 
                         & RFL \cite{schulter2015fast}             & 29.11/0.8904  & 25.86/0.7900  & 24.19/0.7096  \\ \cline{2-5} 
                         & SelfEx \cite{huang2015single}          & 29.54/0.8967  & 26.44/0.8088  & 24.79/0.7374  \\ \cline{2-5} 
                         & RED \cite{mao2016image}             & 30.91/0.9159  & 27.31/0.8303  & 25.35/0.7587  \\ \cline{2-5} 
                         & DnCNN \cite{zhang2012single}           & 30.74/0.9139  & 27.15/0.8276  & 25.20/0.7521  \\ \cline{2-5} 
                         & TNRD \cite{chen2016trainable}            & 29.70/0.8994  & 26.42/0.8076  & 24.61/0.7291  \\ \cline{2-5} 
                         & FDSR \cite{lu2018fast}            & 30.91/0.9088  & 27.23/0.8190  & 25.27/0.7417  \\ \cline{2-5} 
                         & DRCN \cite{kim2016deeply}            & 30.75/0.9133  & 27.15/0.8276  & 25.14/0.7510  \\ \cline{2-5} 
                         & LapSRN \cite{lai2017deep}             & 30.41/0.9100  & -  & 25.21/0.7560  \\ \cline{2-5}
                         & WaveResNet \cite{bae2017beyond}             & 30.96/0.9169 & 27.28/0.8334 & 25.36/0.7614  \\ \cline{2-5}
                         & CPCA \cite{xu2018self}             & 28.17/0.8990 & 25.61/0.8123 & 23.62/0.7257  \\ \cline{2-5}
                         & NDRCN \cite{cao2019new}             & 31.06/0.9175 & 27.23/0.8312 & 25.16/0.7546  \\ \cline{2-5}
                         & MemNet \cite{tai2017memnet}          & 31.31/0.9195  & 27.56/0.8376  & 25.50/0.7630  \\ \cline{2-5} 
                         & LESRCNN \cite{tian2020lightweight}         & 31.45/0.9206  & 27.70/0.8415  & 25.77/0.7732  \\ \cline{2-5} 
                         & LESRCNN-S \cite{tian2020lightweight}       & 31.45/0.9207  & 27.76/0.8424  & 25.78/0.7739  \\ \cline{2-5} 
                         & ScSR \cite{yang2010image}            & 28.26/0.8828  & -             & 24.02/0.7024  \\ \cline{2-5} 
                         & DSRCNN \cite{song2022dual} & \textcolor{blue}{31.83/0.9252}  & \textcolor{blue}{27.99/0.8483}  & \textcolor{blue}{25.94/0.7815} \\ \cline{2-5}
                         & DAN \cite{huang2020unfolding}             & 30.60/0.9060  & 27.65/0.8352  & 25.86/0.7721  \\ \cline{2-5} 
                         & LDRAN\cite{sahambi2023lightweight}            & -             & -             & 25.91/0.7786  \\ \cline{2-5} 
                         & PGAN \cite{shi2023structure}            & -             & -             & 25.47/0.9574  \\ \cline{2-5} 
                         & ADSRNet   (Ours)         & \textcolor{red}{32.00/0.9267}  & \textcolor{red}{28.02/0.8493}  & \textcolor{red}{26.01/0.7827}  \\ \hline
\end{tabular}
}
\end{table}
\begin{table}[!t]
\caption{Running time (millisecond) of different methods for image super-resolution via different sizes for $\times4$.\label{tab:table6}}
\centering
\scalebox{0.91}{
\begin{tabular}{|c|c|c|c|c|}
\hline
Image sizes & VDSR \cite{kim2016accurate} & CARN-M \cite{ahn2018fast} & ACNet \cite{tian2021asymmetric} & ADSRNet   (Ours) \\ \hline
256 × 256  & 17.2         & 15.9           & 18.38          & 14.17            \\ \hline
512 × 512  & 57.5         & 19.9           & 75.79          & 26.11            \\ \hline
\end{tabular}
}
\end{table}
\begin{table}[!t]
\caption{Parameters and flops of different methods with scale factor of 4 for restoration high-quality images of $1024\times1024$.\label{tab:table7}}
\centering
\scalebox{1.1}{
\begin{tabular}{|c|c|c|c|c|}
\hline
Methods        & Parameters & Flops   \\ \hline
DCLS \cite{luo2022deep}  & 13,626K     & 498.18G \\ \hline
ACNet \cite{tian2021asymmetric} & 1,357K      & 132.82G \\ \hline
ADSRNet (Ours) & 1,819K      & 110.99G \\ \hline
\end{tabular}
}
\end{table}
\begin{figure*}[!t]
\centering
\includegraphics[width=7.0in]{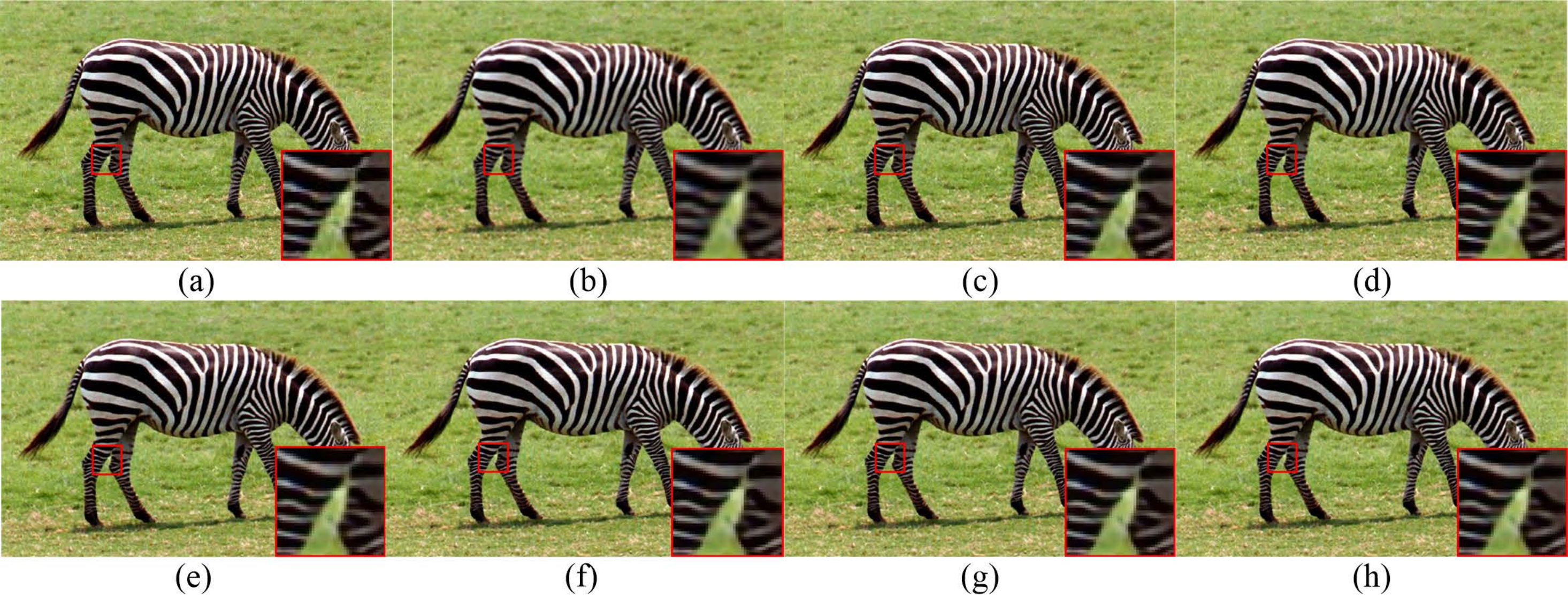}
\caption{Predicted high-quality images from different methods on a same low-resolution image from Set14 for ×2: (a) A HR image, (b)Bicubic, (c) SRCNN, (d) LESRCNN, (e) DCLS, (f) VDSR, (g) DRCN and (h) ADSRNet (Ours).}
\label{fig_2}
\end{figure*}
\begin{figure*}[!t]
\centering
\includegraphics[width=7.0in]{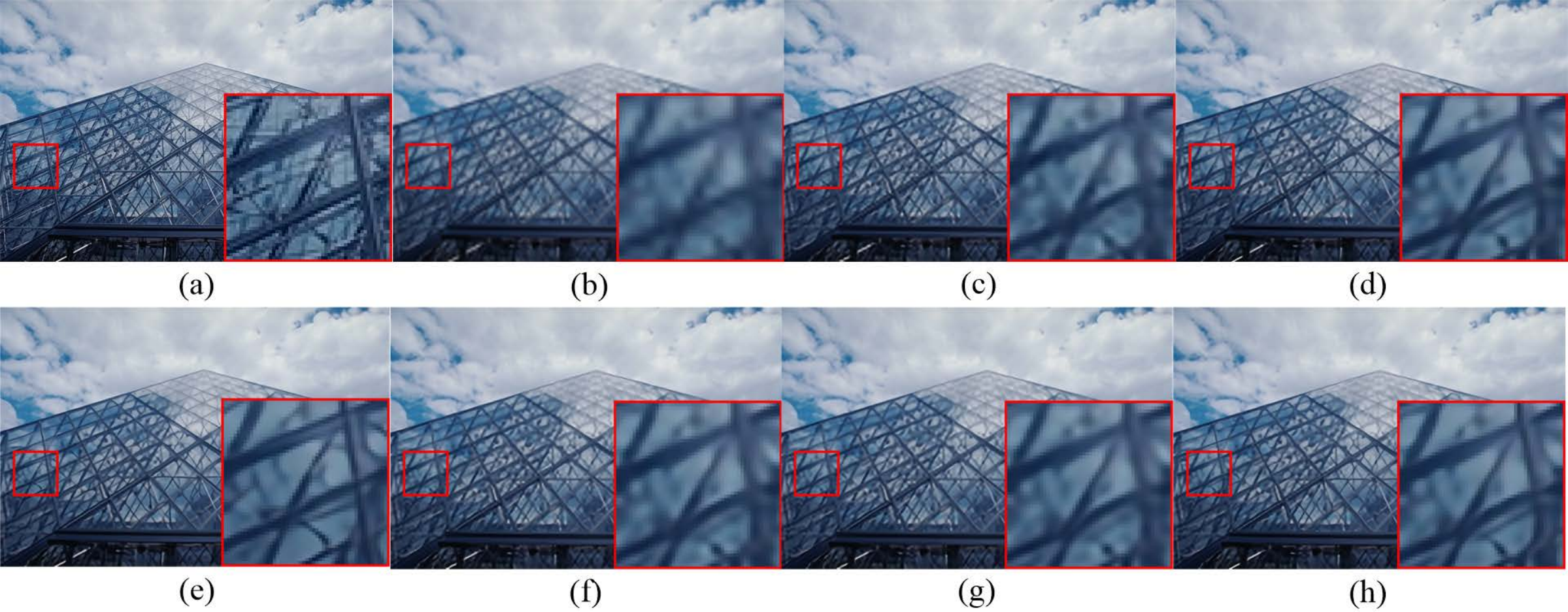}
\caption{Predicted high-quality images from different methods on a same low-resolution image from B100 for ×3: (a) A HR image, (b)Bicubic, (c) SRCNN, (d) LESRCNN, (e) DCLS, (f) VDSR, (g) DRCN and (h) ADSRNet (Ours).}
\label{fig_3}
\end{figure*}
\begin{figure*}[!t]
\centering
\includegraphics[width=7.0in]{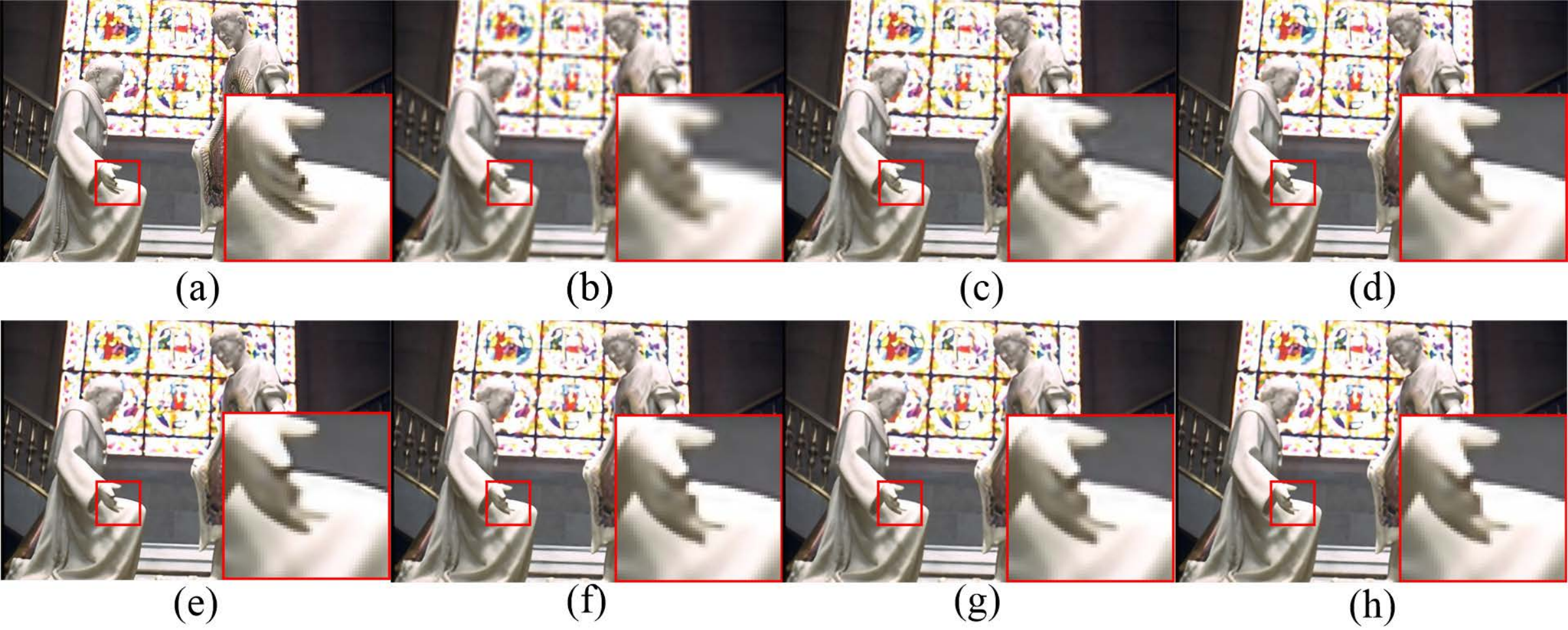}
\caption{Predicted high-quality images from different methods on a same low-resolution image from B100 for ×3: (a) A HR image, (b)Bicubic, (c) SRCNN, (d) LESRCNN, (e) DCLS, (f) VDSR, (g) DRCN and (h) ADSRNet (Ours).}
\label{fig_4}
\end{figure*}
\begin{figure*}[!t]
\centering
\includegraphics[width=7.0in]{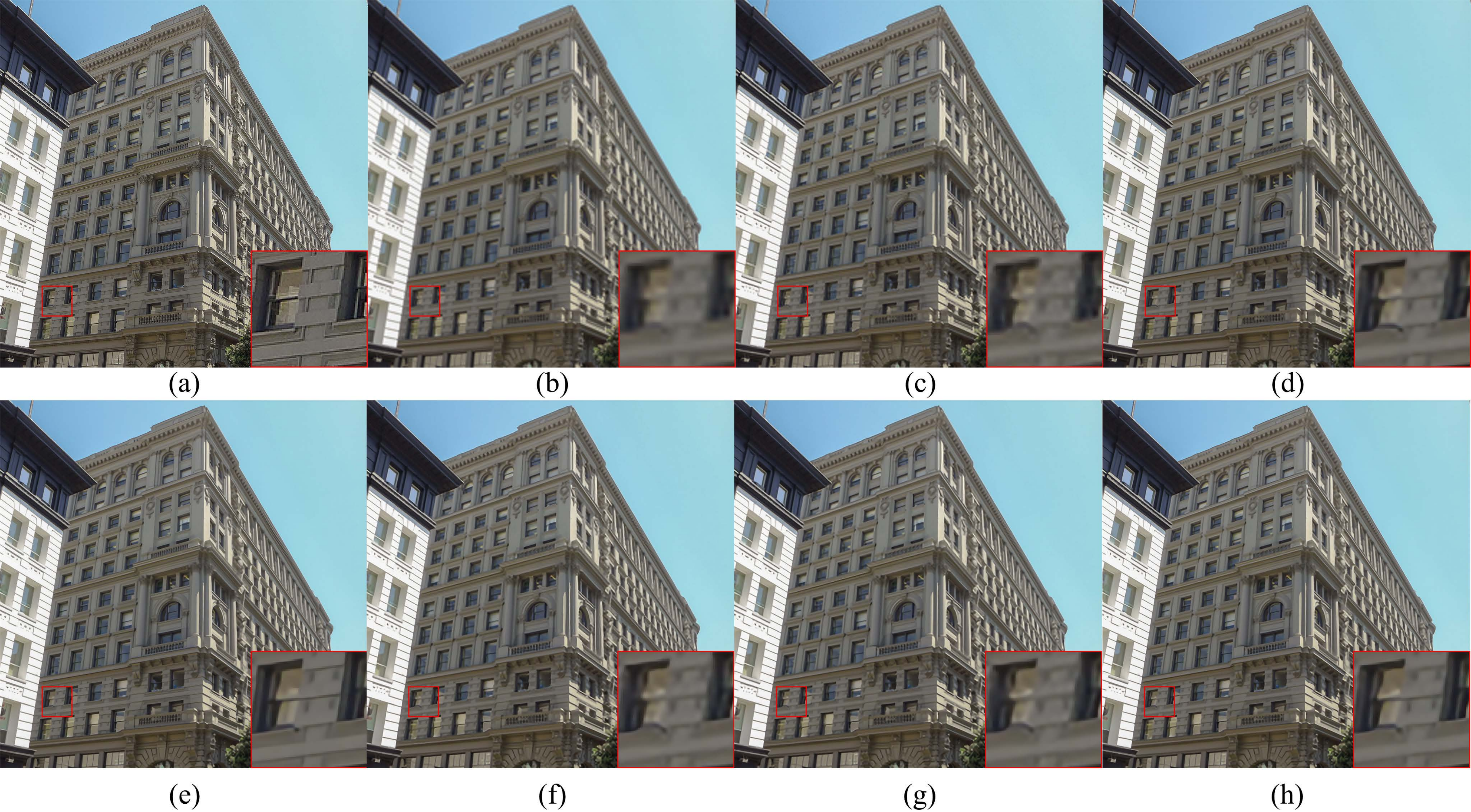}
\caption{Predicted high-quality images from different methods on a same low-resolution image from U100 for ×4: (a) A HR image, (b)Bicubic, (c) SRCNN, (d) LESRCNN, (e) DCLS, (f) VDSR, (g) DRCN and (h) ADSRNet (Ours).}
\label{fig_5}
\end{figure*}
\begin{figure*}[!t]
\centering
\includegraphics[width=7.0in]{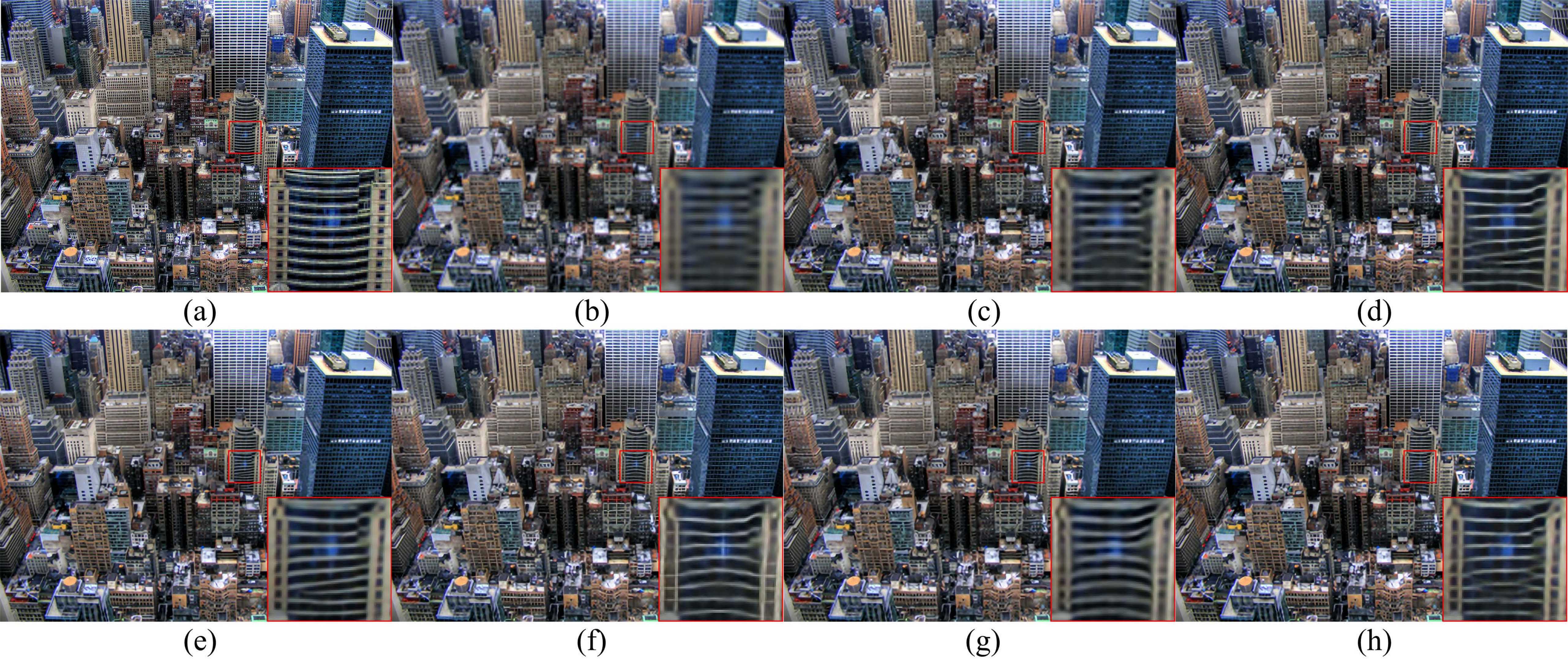}
\caption{Predicted high-quality images from different methods on a same low-resolution image from U100 for ×4: (a) A HR image, (b)Bicubic, (c) SRCNN, (d) LESRCNN, (e) DCLS, (f) VDSR, (g) DRCN and (h) ADSRNet (Ours).}
\label{fig_6}
\end{figure*}
\begin{figure*}[!t]
\centering
\includegraphics[width=7.0in]{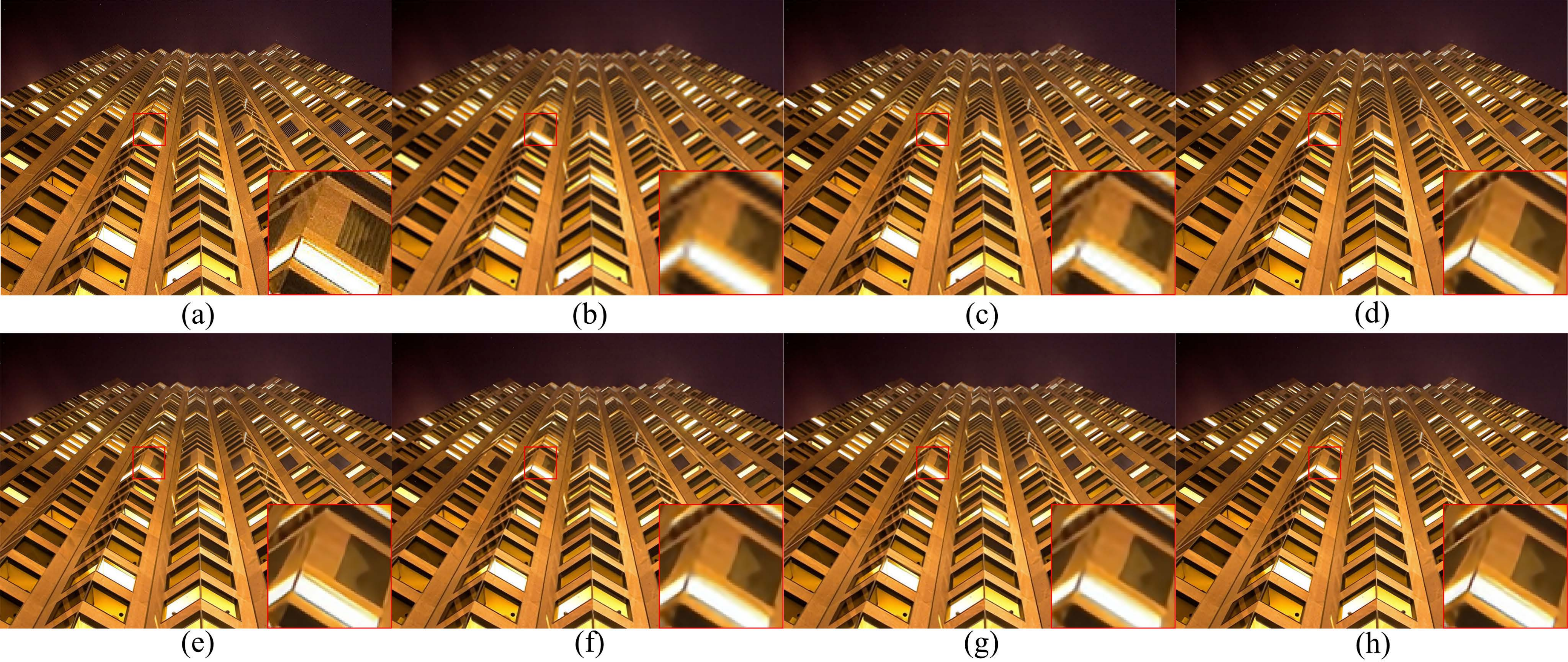}
\caption{Predicted high-quality images from different methods on a same low-resolution image from U100 for ×4: (a) A HR image, (b)Bicubic, (c) SRCNN, (d) LESRCNN, (e) DCLS, (f) VDSR, (g) DRCN and (h) ADSRNet (Ours).}
\label{fig_7}
\end{figure*}

The first phase design a 16-layer stacked CRU to extract low-frequency structural information, according to VGG \cite{simonyan2014very}. Its effectiveness can be proved via ‘Heterogeneous upper sub-network and a CB’ and ‘ADSRNet without RL operations in symmetrical lower sub-network’ in TABLE I. To enhance relationship of hierarchical features, the second phase is conducted. It uses RL operations to gather obtained features of shallow and deep layers to achieve a symmetrical lower sub-network to extract more accurate information as shown in Section III.D. As illustrated in TABLE I, we can see that ‘ADSRNet’ has obtained higher PSNR and SSIM than that of ADSRNet without RL operations in symmetrical lower sub-network, which shows effectiveness of RL operations in the symmetrical lower sub-network for image SR. Besides, 'ADSRNet' has obtained higher PSNR value than that of 'Heterogeneous upper sub-network and a CB' in TABLE I, which shows effectiveness of Heterogeneous parallel networks for image super-resolution.

Construction block: To construct HR images, we design a 2-layer construction block. It consists of a sub-pixel convolutional layer and a convolutional layer. Sub-pixel convolutional layer is used to amplify low-frequency features to high-frequency features. A convolutional layer is utilized to construct predicted HR images. 
\subsection{Comparisons with Popular Methods for SR}
A heterogeneous network architecture can more diversified structural information, which can provide complementary information to strengthen robustness of an obtained SISR model. The heterogeneous network is deigned, according to enhancing relation of context information, salient information relation of a kernel mapping and relations of shallow and deep layers. Besides, a deep network via parallel networks can extract extra structural information to facilitate more representative information to further improve SR effect.
These make our ADSRNet obtain better SR results than these of some classical SR methods, i.e., VDSR \cite{kim2016accurate}, DRCN \cite{kim2016deeply}. Rationality and effectiveness of key techniques in our ADSRNet are given in Section IV. D. Superiority of the proposed ADSRNet for SISR is given as follows.
In this section, we use both quantitative and qualitative analysis to evaluate results of our ADSRNet on SISR. Quantitative analysis includes PSNR, SSIM, running time and complexity of our ADSRNet. Specifically, PSNR and SSIM are used to measure quality of predicted HR images. Also, running time and parameters are used to test feasibility of our ADSRNet for real applications, i.e., phones and cameras. We use A+ \cite{timofte2015a+}, jointly optimized regressors (JOR) \cite{dai2015jointly}, image upscaling with super-resolution forests (RFL) \cite{schulter2015fast}, self-exemplars SR method (SelfEx) \cite{huang2015single}, cascade of sparse coding based network (CSCN) \cite{wang2015deep}, image restoration using encoder-decoder networks (RED) \cite{mao2016image}, denoising convolutional neural network (DnCNN) \cite{zhang2012single}, trainable non-linear reaction diffusion (TNRD) \cite{chen2016trainable}, fast dilated residual SR convolutional net-work (FDSR) \cite{lu2018fast}, SRCNN \cite{dong2015image}, FSRCNN \cite{dong2016accelerating}, residue context sub-network (RCN) \cite{shi2017structure}, VDSR \cite{kim2016accurate}, DRCN \cite{kim2016deeply}, 
IDN \cite{hui2018fast}, DRRN \cite{tai2017image}, Laplacian SR network (LapSRN) \cite{lai2017deep}, 
new architecture of deep recursive convolution networks for SR (NDRCN) \cite{cao2019new}, a persistent memory network (MemNet) \cite{tai2017memnet}, CARN-M \cite{ahn2018fast}, light-weight image super-resolution with enhanced CNN (LESRCNN) \cite{tian2020lightweight}, deep alternating network (DAN) \cite{huang2020unfolding}, 
Pixel-Level Generative Adversarial Network (PGAN) \cite{shi2023structure} and our ADSRNet on four public datasets, i.e., Set5, Set14, B100 and U100 for x2, x3 and x4 to conduct experiments. As shown in TABLEs II and III, we can see that our ADSRNet has obtained the best performance in terms of PSNR and SSIM for x2, x3 and x4. For instance, our ADSRNet has improvements of 0.12dB on PSNR and 0.004 on SSIM on Set 5 for x2 than that of IDN in TABLE II. Our ADSRNet has obtained improvements of 0.1dB on PSNR and 0.003 on SSIM on Set14 for x4 than that of DSRCNN in TABLE III. That also shows that our ADSRNet is effective on small datasets for image super-resolution. For big datasets, our ADSRNet still has an advantage for image super-resolution in TABLEs IV and V, we can see that our ADSRNet has obtained the best SR results for all the scale factors, i.e., x2, x3 and x4. For instance, our ADSRNet has exceeded 0.09dB on PSNR and 0.0016 on SSIM than that of DSRCNN for x2 on B100 in TABLE IV. Our ADSRNet has exceeded 0.07dB on PSNR and 0.0012 on SSIM than that of DSRCNN for x4 on U100 in TABLE V. That shows that our ADSRNet is suitable to big datasets for image super-resolution. Specifically, red and blue lines denote the best and second SR results from TABLE II to TABLE V, respectively.

To test practicality of our ADSRNet, we use running time and complexity to test performance of our ADSRNet for image SR. As shown in TABLE VI, our ADSRNet has obtained competitive running time for restoring LR images sizes of $256\times256$ and $512\times512$. As illustrated in TABLE VII, although our ADSRNet has obtained more parameters than that of ACNet, it has obtained less flops than that of ACNet. Thus, it is competitive in complexity and suitable for application in consumer electronic products. 
According to experiment results, our ADSRNet is effective in terms of quantitative analysis.

For qualitative analysis, we use Bicubic, SRCNN, LESRCNN, DCLS, VDSR, DRCN and ADSRNet on a low-resolution image from the B100 and U100 for x3 and x4 to recover high-quality images, which are used to compare with given HR images. That is, we amplify one area of predicted high-quality images from different methods as observation areas, observation areas are clearer, their corresponding methods are more effective for SISR. As shown in Figs.2-7, our ADSRNet has obtained clearer areas, it shows that our ADSRNet is effective for qualitative analysis. In a summary, our ADSRNet is a good tool for image resolution, according to quantitative and qualitative analysis.
\section{Conclusion}
In this paper, we propose a adaptive convolutional network via a parallel network in SISR. The upper network depends on enhancing relation of context information, salient information relation of a kernel mapping and relations of shallow and deep layers to implement efficient SR performance. 
The lower network relies on a symmetric architecture to strengthen relation of 
hierarchical information to improve effect of SISR. Two networks can extract complementary information to facilitate more robust information to protect performance of SISR for different scenes. We will deal with SISR with non-reference images in the future.





\ifCLASSOPTIONcaptionsoff
  \newpage
\fi



%
\bibliographystyle{IEEEtran}
\bibliography{IMSC_AGL}

\end{document}